\documentclass[epj]{svjour}
\usepackage{graphics}
\begin{document}
\title{Nucleon resonance decay by the $K^0\Sigma^+$ channel}
\author{
R. Castelijns
\inst{1,10},
A.V. Anisovich
\inst{2,3},
G. Anton
\inst{4},
J.C.S. Bacelar
\inst{1},
B. Bantes
\inst{5},
O. Bartholomy
\inst{2},
D. Bayadilov
\inst{2,3},
Y.A. Beloglazov
\inst{3},
R. Bogend\"orfer
\inst{4},
V. Crede
\inst{2,11},
H. Dutz
\inst{5},
A. Ehmanns
\inst{2},
D. Elsner
\inst{5},
K. Essig
\inst{2},
R. Ewald
\inst{5},
I. Fabry
\inst{2},
H. Flemming
\inst{6},
K. Fornet-Ponse
\inst{5},
M. Fuchs
\inst{2},
C. Funke
\inst{2},
R. Gothe
\inst{5,12},
R. Gregor
\inst{7},
A.B. Gridnev
\inst{3},
E. Gutz
\inst{2},
S. H\"offgen
\inst{5},
P. Hoffmeister
\inst{2},
I. Horn
\inst{2},
J. H\"ossl
\inst{4},
I. Jaegle
\inst{8},
J. Junkersfeld
\inst{2},
H. Kalinowsky
\inst{2},
S. Kammer
\inst{5},
Frank Klein
\inst{5},
Friedrich Klein
\inst{5},
E. Klempt
\inst{2},
H. Koch
\inst{6},
M. Konrad
\inst{5},
B. Kopf
\inst{6},
M. Kotulla
\inst{7,8},
B. Krusche
\inst{8},
J. Langheinrich
\inst{5,12},
H. L\"ohner
\inst{1},
I.V. Lopatin
\inst{3},
J. Lotz
\inst{2},
S. Lugert
\inst{7},
H. Matth\"ay
\inst{6},
D. Menze
\inst{5},
T. Mertens
\inst{8},
J.G. Messchendorp
\inst{1},
V. Metag
\inst{7},
C. Morales
\inst{5},
M. Nanova
\inst{7},
V.A. Nikonov
\inst{2,3},
D. Novinski
\inst{3},
R. Novotny
\inst{7},
M. Ostrick
\inst{5,9},
L.M. Pant
\inst{7,13},
H. van Pee
\inst{2,7},
M. Pfeiffer
\inst{7},
A. Roy
\inst{7,14},
A. Radkov
\inst{3},
A.V. Sarantsev
\inst{2,3},
S. Schadmand
\inst{7,10},
C. Schmidt
\inst{2},
H. Schmieden
\inst{5},
B. Schoch
\inst{5},
S. Shende
\inst{1},
A. S\"ule
\inst{5},
G. Suft
\inst{4},
V.V. Sumachev
\inst{3},
T. Szczepanek
\inst{2},
U. Thoma
\inst{2,7},
D. Trnka
\inst{7},
R. Varma
\inst{7,14},
D. Walther
\inst{5},
C. Weinheimer
\inst{2,15},
 \and
C. Wendel
\inst{2}
\newline
(The CBELSA/TAPS Collaboration)
}
\mail{ loehner@kvi.nl}

\institute{
KVI, University of Groningen, The Netherlands \and
Helmholtz-Institut f\"ur Strahlen- und Kernphysik, Universit\"at Bonn, Germany \and
Petersburg Nuclear Physics Institute, Gatchina, Russia \and
Physikalisches Institut, Universit\"at Erlangen, Germany \and
Physikalisches Institut, Universit\"at Bonn, Germany \and
Physikalisches Institut, Ruhr-Universit\"at Bochum, Germany \and
II. Physikalisches Institut, Universit\"at Gie\ss en, Germany \and
Physikalisches Institut, Universit\"at Basel, Switzerland \and
Institut f\"ur Kernphysik, Universit\"at Mainz, Germany \and
\emph{Present address:} Institut f\"ur Kernphysik, FZ J\"ulich, Germany \and
\emph{Present address:} Florida State University, FL, USA \and
\emph{Present address:} University of South Carolina, SC, USA \and
\emph{On leave from:} Nuclear Physics Division, BARC, Mumbai, India \and
\emph{On leave from:} Department of Physics, I.I.T. Powai, Mumbai, India \and
\emph{Present address:} Institut f\"ur Kernphysik, Universit\"at M\"unster, Germany
}

%
%
%
\date{Received: date / Revised version: date}
%
\abstract
{
Hyperon production off the proton in the $K^0 \Sigma^+$ channel has been studied at the tagged photon beam facility at the ELSA electron accelerator in Bonn. This experiment was part of a series of neutral meson production experiments on various targets. For this purpose, the Crystal Barrel and TAPS photon spectrometers have been combined to provide a $4 \pi$ detector for multi-neutral-particle final states. A high-quality excitation function, recoil polarizations, and angular distributions from  threshold up to 2.3 GeV center-of-mass energy were obtained. The results are compared with predictions of recent coupled-channels calculations within the K-matrix formalism and are interpreted by the partial-wave analysis within the Bonn-Gatchina model.
}      
\titlerunning{Nucleon resonance decay by the $K^0\Sigma^+$ channel}
\authorrunning{R. Castelijns et al.}

\maketitle
\section{\label{sec:introduction}Introduction}
Since the ground state octet and decuplet of baryons were discovered, efforts have been underway to explain the
structure of the spectrum of their excited states. Experimentally many baryon resonances have been found,
most of them in pion elastic scattering experiments. It is now clear that the spectrum of these excited states consists of
many states, each with a large width of up to a few hundred MeV. Since the states overlap and interfere with
each other, the interpretation of experimental data and the extraction of resonance parameters is extremely difficult.
Although the theory describing the interactions between the quarks, QCD, is known, the perturbative methods commonly
used to treat the QCD Lagrangian are not applicable to calculate the resonance properties. The running
coupling constant of the strong interaction is no longer small compared to unity at the relatively low
quark energies involved in baryon resonances.

To still gain insight into the structure of the resonance spectrum, even though it cannot be deduced from first principles,
QCD based models \cite{ca,gl,lo,lo2} have been developed. These models are
able to explain the mass spectrum of the experimentally observed baryon resonances reasonably well, but all predict many
more states than have been observed experimentally. This discrepancy may either be explained by an uncertainty in the
number of degrees of freedom used by the models or by assuming the unobserved resonances not to couple strongly to
the $N \pi$ channel in which most of the resonance data so far have been gathered.

Recent predictions in a quark-pair creation model \cite{ca2} or a collective string-like three-quark
model \cite{le} revealed substantial decay branches of baryon resonances into the $K\Lambda$ and
$K\Sigma$ channels. Kaon production experiments will therefore be an important tool to establish
or disprove ``missing'' resonances and thus to determine the relevant degrees of freedom of quark
models.
To interpret the experimental data it is important to treat all production and decay channels on an
equal footing, using a coupled-channel analysis. Therefore it is important to obtain accurate cross
sections including the full angular distributions and the polarization observables.

\section{\label{sec:experiment}Experiment}
Here we present the measurement of such data for the photoproduction of $K^0$ mesons off the proton.
The associated strangeness production requires a hyperon ($\Sigma^+$) in the final state.
The neutral decay channel has been exploited to identify the reaction:
\begin{equation}
\gamma p \rightarrow K^0 \Sigma^+ \rightarrow (\pi^0 \pi^0) (\pi^0 p) \rightarrow 6 \gamma p
\end{equation}

This measurement is complementary to earlier measurements \cite{la,ca3} which identify the reaction via the
charged decay of the $K^0$ and the $\Sigma^+$. The advantage of the neutral decay channel lies in the very flat
acceptance of the measurement, covering the entire phase space available to the reaction. In addition, this
method allows to determine the background with high precision, and an accurate relative normalization is provided by
the $\eta$ channel:
\begin{equation}
\gamma p \rightarrow \eta p \rightarrow 3 \pi^0 p \rightarrow 6 \gamma p
\end{equation}

The measurement was performed using the combination of the TAPS \cite{taps} and Crystal Barrel \cite{crystalbarrel}
calorimeters, which
has been set up at the bremsstrahlung-tagger photon beam facility at the ELSA \cite{elsa} electron accelerator
in Bonn.
At this facility, unpolarized and linearly polarized photon beams were created in the energy range from
0.5 to 2.9 GeV by passing the electron
beam through a thin copper radiator or diamond crystal \cite{da}, respectively. The electrons were momentum
analyzed by a combination of a dipole magnet and a tagger detector, consisting of 480 scintillating fibers and 14
scintillator bars in a partly overlapping configuration, to obtain the energy of the emitted bremsstrahlung
photon.

The resulting photon beam with typical intensity of 10$^7$ photons/s was traversing a 5 cm long liquid hydrogen
target, surrounded by the calorimeter combination formed by Crystal Barrel and TAPS.
The Crystal Barrel consisted of 1290 $CsI$ crystals in a cylindrical arrangement providing 16 radiation lengths
and covering
all polar angles from $30^{\circ}$ to $168^{\circ}$, with full azimuthal coverage. The
forward angles, from $5^{\circ}$ to $30^{\circ}$, were covered by TAPS. This calorimeter consisted
of 528 $BaF_2$ crystals with hexagonal cross section and a depth of 12 radiation lengths. TAPS was configured as an hexagonal
wall, thus serving as the forward endcap of the Crystal Barrel cylinder.
Combined, the two calorimeters provided a geometrical acceptance of almost $4\pi$. The
high granularity and good acceptance of this system make it very well suited to
measure reactions with a high number of photons in the final state. The typical invariant mass resolution (rms)
of $\pi^0$ was 5.8\%.

Charged particles emitted in the forward direction were identified online by 5 mm thin plastic scintillators
mounted directly in front of each of the TAPS crystals. Charged particles detected by the Crystal Barrel
were identified via a cylindrical scintillating fiber detector \cite{scifi-erlangen} surrounding the target.
Additionally, an offline analysis of the pulse shape and time-of-flight information provided by TAPS was used
to distinguish the particles emitted in the acceptance of TAPS.

\section{\label{sec:analysis}Analysis}
From the complete data set, gathered in 1400 hours of photon beam on the hydrogen target, 
events containing $3\pi^0$'s and one additional hit
were selected, using windows on three mutually exclusive invariant mass spectra of $2\gamma$ combinations.

To improve the invariant mass resolutions and identify the proton among the seven recorded hits,
each event was kinematically fitted. The constraints used in the kinematic fit were the
conservation of momentum and energy (4 constraints) and the $\pi^0$ invariant masses (3 constraints).
It is important to note that neither the $K^0$ nor the $\Sigma^+$ mass was used as a constraint.
The measurements allowed to vary within their respective resolution are the energies deposited by the six
photons and the two angles of each of their trajectories.
For the proton only the two angles of its trajectory were allowed to vary. Its energy was calculated
from the other observables since protons with a kinetic energy in excess of 400 MeV are not stopped within
the calorimeters. This will result in a large inaccuracy in the measurement of the proton kinetic energy. The 
resulting fit is therefore 6 fold over-constrained.

Identification of the proton was achieved by fitting each event seven times, once for each of the
seven proton candidates. The combination yielding the highest confidence level was taken as the correct
one. As a further condition, the measurement of a particle in TAPS and an electron
in the tagger should be coincident in time. Since the time-of-flight of slow protons was much longer than that of
photons, only the photons identified by the TAPS detector were considered for this coincidence.

Applying the kinematic fit procedure improves the invariant mass resolution in the
$\eta \rightarrow 3 \pi^0$ channel from 3.8\% to 1.5\% while the $\eta$ yield remains constant.
The $K^0$ and the $\Sigma^+$ invariant-mass resolutions improved by a factor 3.
This is particularly important since, due to the conservation of
strangeness in the strong decay, both particles have to decay weakly and therefore have a relatively long
lifetime. Both particles decay after traveling on average 4 cm, broadening the measured
invariant mass distribution of the $K^0$ approximately by a factor 2. Events for which the kinematic fit yielded a confidence level
of less than 10\% were removed from the data to reduce background.

\begin{figure}
\resizebox{0.50\textwidth}{!}{%
  \includegraphics{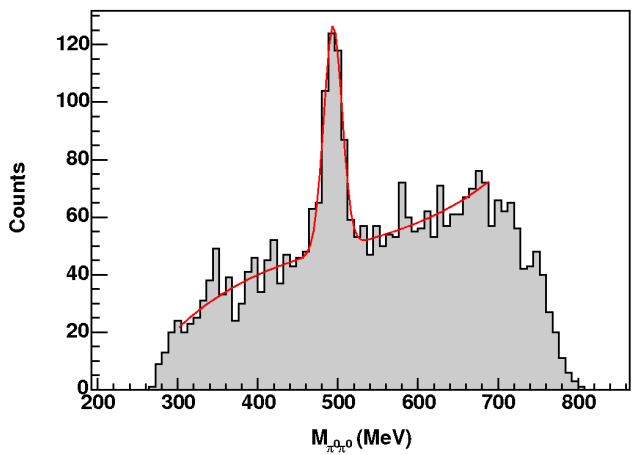}
}
\caption{\label{figkaon}The $\pi^0 \pi^0$ invariant mass spectrum after the kinematic
fit, and gated on the $\Sigma^+$ mass in the $p \pi^0$ invariant mass spectrum, for incoming
photon energies between 1450 and 1550 MeV, integrated over all angles.
The line shows the combined fit of the background and the $K^0$ signal.}
\end{figure}

Those events in which the three $\pi^0$'s are produced via decay of an $\eta$ meson
were removed from the data set by selecting the region outside the $\eta$ peak in the $3\pi^0$ invariant
mass spectrum. At this point, the remaining background was mainly caused by $3\pi^0$
sequential resonance decay and, to a lesser degree, combinatorics. Background was further
reduced by removing all events that did not contain a $\Sigma^+$ hyperon. This condition was achieved by
selecting the region within the $\Sigma^+$ invariant mass peak in the $p \pi^0$ invariant
mass spectrum. An example of the $\pi^0 \pi^0$ invariant mass of the remaining events is shown as histogram
in figure~\ref{figkaon}.
In order to achieve a consistent description of the background in different kinematical regions, it was
fitted by a fourth-order polynomial. Figure~\ref{figkaon} also shows the combination of the polynomial
and a Gaussian fit to the  $K^0$ signal. For the complete data set we obtained background-fit parameters
which are smoothly varying functions of the center-of-mass angle and the
incident photon energy.

In total a number of 10000 $K^0$'s was obtained. The $K^{0}_{long}$ cannot contribute significantly to this yield, since the decay
$K^{0}_{long} \rightarrow 2 \pi^0$ is suppressed by three orders of magnitude due to CP violation \cite{PDG}.

The acceptance for the process $\gamma p \rightarrow K^0 \Sigma^+ \rightarrow 3 \pi^0 p \rightarrow 6 \gamma p$
was determined using a Monte Carlo simulation. The simulated events were distributed evenly over the available phase space.
To calculate the acceptance the simulated data set was analyzed using the same event selection criteria and kinematical
fitting procedures as were used to analyze the measured data. In addition, the simulated data were biased by the same threshold
conditions as were imposed by the experimental trigger.
The resulting acceptance covers all
center-of-mass angles and varies only little, between 11\% and 14\%,
with the incoming photon energy and the center-of-mass angle of the reaction.
This insensitivity of the acceptance is caused by the decay of the two primary particles, $K^0$ and $\Sigma^+$,
and is very advantageous since it ensures a reliable determination of the acceptance.
The losses of long-living $K^0$ and $\Sigma^+$ were well reproduced by the Monte Carlo simulations.

The acceptance for the process $\gamma p \rightarrow \eta p \rightarrow 3\pi^0 p \rightarrow 6\gamma p$ was
obtained using the same procedure. This reaction was concurrently measured with the reaction of interest and
therefore provided an extra check on the acceptance calculation and the
determination of the experimental luminosity. The cross sections for $\eta$ production were well
determined and published recently \cite{cr}. The energy dependence and the shape of the angular distributions
for $\eta$ production obtained from this work was therefore compared with the previously published data.
The good agreement between the angular distributions of both data sets confirms the acceptance calculation.
 
The $\gamma p \rightarrow \eta p$ differential cross sections have also been used to verify the absolute
normalization of the cross sections for the reaction $\gamma p \rightarrow K^0\Sigma^+$.
The energy dependence of the derived photon flux agrees well with the experimental determination and the
theoretical description of the bremsstrahlung spectrum \cite{ya} including QED corrections.

\section{\label{sec:results}Results}
\begin{figure}
\resizebox{0.55\textwidth}{!}{%
  \includegraphics{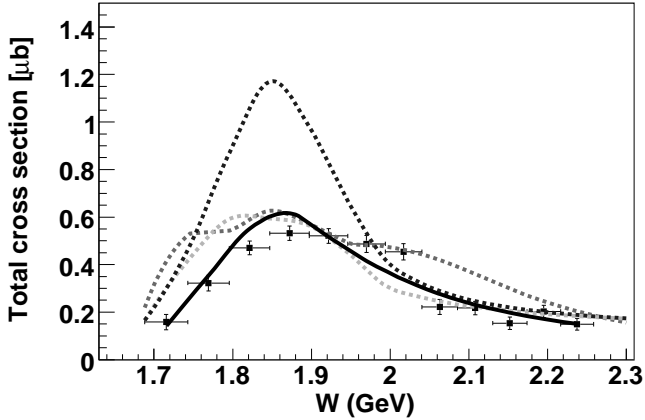}
}
\caption{\label{figextheo}The excitation function for $K^0\Sigma^+$ photoproduction measured by this
experiment, compared with the predictions by the Bonn-Gatchina model \protect\cite{an,sa2} (solid line) and
the Usov-Scholten model \protect\cite{us} (dashed lines). The different dashed lines represent the results of the model 
including all established resonances (black), an additional $P_{13}(1830)$ resonance (light gray),
and the result of a full $\chi^2$ minimization (dark gray, preliminary result) over the data in all available channels
with the additional resonance and extra contact terms.}
\end{figure}

\begin{figure}
\resizebox{0.50\textwidth}{!}{%
  \includegraphics{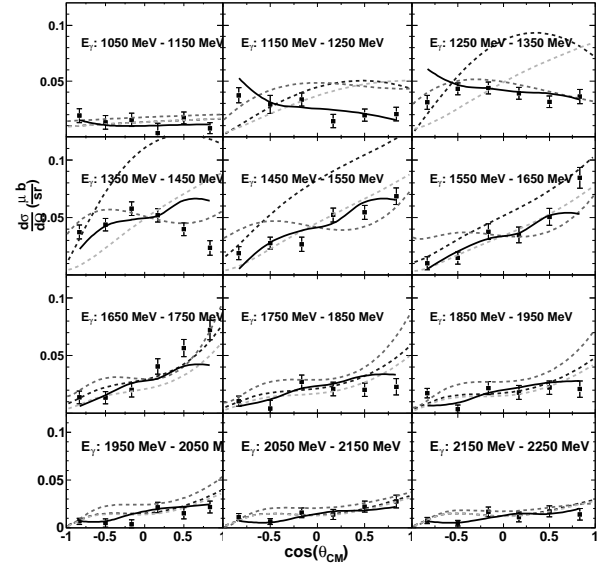}
}
\caption{\label{figdiftheo}The differential cross sections for $K^0\Sigma^+$ photoproduction measured by this experiment and compared with the predictions (lines, see caption figure~\ref{figextheo}).}
\end{figure}

\begin{figure}
\resizebox{0.50\textwidth}{!}{%
  \includegraphics{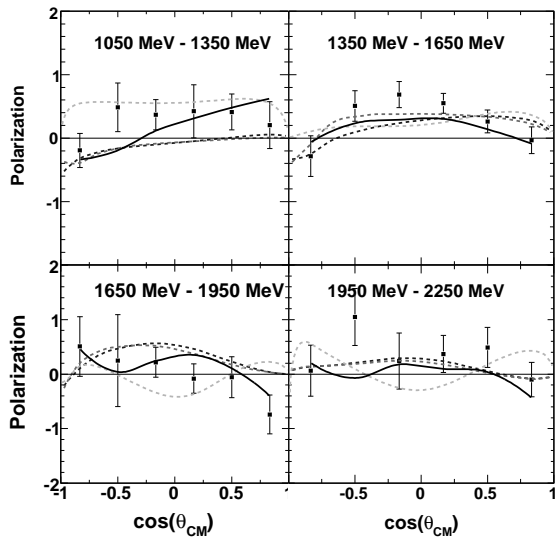}
}
\caption{\label{figpoltheo}The recoil polarization for the $\Sigma^+$ hyperon measured by this experiment and
compared with the predictions (lines, see caption figure~\ref{figextheo}).}
\end{figure}

The excitation function measured in this experiment \cite{me} is shown in figure~\ref{figextheo}. In addition,
the differential cross sections and the recoil polarization are shown in figures~\ref{figdiftheo}
and \ref{figpoltheo}, respectively.

\begin{figure}
\resizebox{0.55\textwidth}{!}{%
  \includegraphics{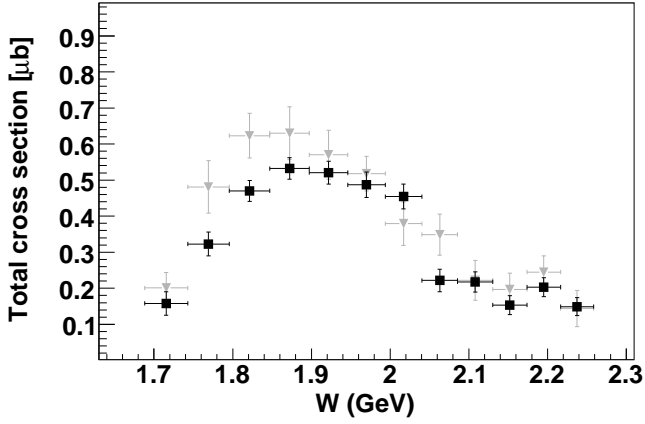}
}
\caption{\label{figexexp}The excitation function for $K^0\Sigma^+$ photoproduction measured by this
experiment (black squares), compared with the results of the SAPHIR experiment \protect\cite{la} (gray triangles)}
\end{figure}

The $K^0 \Sigma^+$ excitation function and differential cross sections measured in the present work largely
agree with the results published recently by the SAPHIR collaboration \cite{la}, shown in figure~\ref{figexexp}. 
For the energies near threshold,
our results do not reveal
the peak at forward angles in the angular distribution shown by the SAPHIR result, and therefore the excitation
function is somewhat lower in that region. Good agreement is also obtained with the CLAS data \cite{ca3}.

Due to the strange quark content of the $\Sigma^+$ hyperon, the decay has to proceed via the weak interaction.
Because the decay is parity violating, the hyperon polarization $P_{y}$ can be determined experimentally.
From the decay asymmetry of the $\Sigma^+$ with respect to the reaction plane, the polarization is determined
by averaging the proton angular distribution in the upper and lower half of the reaction plane, resulting in

\begin{equation}
 P = \frac{2}{\alpha} \frac{N_{up} - N_{down}}{N_{up} + N_{down}}
\end{equation}

where $\alpha$ is the decay parameter for $\Sigma^+ \rightarrow \pi^0 p$ and has the
value \cite{PDG} $-0.980 \pm 0.016$.
$N_{up}$ and $N_{down}$ represent the number of events with the decay proton emitted above and below the
reaction plane, respectively.
The resulting hyperon polarization is shown in figure~\ref{figpoltheo} as a function of the
cosine of the center-of-mass angle.

In figure~\ref{figextheo} the excitation function is compared to the full coupled-channels K-matrix calculations of Usov and Scholten \cite{us}.
Taking all established resonances into account, the peak in the total cross section is predicted too high by almost a factor 2.
When taking an additional $P_{13}(1830)$ into account, much better agreement is obtained while simultaneously maintaining
a good description of the data in the $K^+ \Sigma^0$ channel. This reduction of the cross section is caused
by destructive interference effects between the newly assumed resonance and the $P_{33}(1855)$ already present in the model.

Although the inclusion of the $P_{13}(1830)$ greatly improves the agreement with data, discrepancies still remain. 
It is premature to conclude on the basis of the present 
calculations that there is strong evidence for a 
$P_{13}(1830)$ resonance. The observed effects on the cross 
sections could also be generated by additional contact 
terms and further investigations are ongoing.

To investigate the influence of a third $S_{11}$ resonance at 1729 MeV as proposed by Saghai \cite{sa} the
data have been compared with the
predictions obtained from the K-matrix calculations \cite{us}, using two sets of input
parameters. The sets were obtained by fitting the $\eta$ photoproduction cross section from the Crystal
Barrel \cite{cr}
once with and once without the proposed third $S_{11}$. The parameters describing the $S_{11}(1535)$
and the $S_{11}(1650)$ were not refitted as they are determined by the $\pi$ sector.
Both sets of parameters describe the eta photoproduction data well, but the effect of the inclusion of the 
additional $S_{11}$ resonance has only a minor effect on the $K^{0} \Sigma^{+}$ photoproduction cross section, well below the 
experimental sensitivity. The recoil polarization is also not sensitive to the effect of the additional $S_{11}$ 
resonance.

The solid line in figures~\ref{figextheo}-\ref{figpoltheo} shows the prediction of the Bonn-Gatchina model \cite{an,sa2}. The model describes virtually all
existing photoproduction data, including the recent high-precision data from ELSA \cite{ba,cr,gl,la}, GRAAL \cite{aj,ba2}, JLab \cite{mc,ca}, MAMI \cite{kr},
and SPring-8 \cite{ze}. The predicted distributions are in very good agreement with the data; no re-adjustment of parameters was needed.
Both the Usov-Scholten and the Bonn-Gatchina model require a new baryon resonance (not yet listed in the Particle Data Listings \cite{PDG})
at about 1840 MeV to describe the data. Usov and Scholten suggest a $P_{13}(1830)$ resonance, a new $P_{33}(1855)$ had already been 
introduced in their model. The Bonn-Gatchina model requires a new $P_{11}(1840)$.

The Bonn-Gatchina result has been tested by replacing the $P_{11}(1840)$ resonance by a $P_{13}(1830)$ and a $P_{33}(1855)$, both with 
adjustable mass, widths, and helicity couplings. The fit had a significantly smaller likelihood. The data were compatible with a $P_{11}(1840)$
describing photoproduction of $N \pi$, $p \eta$ and $K^+ \Lambda$, but decoupling from the $K \Sigma$ channels, and a $P_{13}(1830)$ and
$P_{33}(1855)$ decaying strongly into $K \Sigma$ but at most very weakly to all other channels. Compared to \cite{us}, 10 more parameters 
are introduced in this ansatz and the same likelihood is achieved. The strong coupling of $P_{13}(1830)$ to $K \Sigma$ in absence of
other couplings would make the $P_{13}(1830)$ a candidate for an exotic state. The alternative solution with a $P_{11}(1840)$ does not
require resonances with anomalous properties.

\section{\label{sec:conclusions}Conclusions}
This analysis of the reaction $\gamma p \rightarrow K^0 \Sigma^+$ produced differential cross sections,
$\Sigma^+$ recoil polarizations, and the excitation function in the region between threshold (1050 MeV)
and 2250 MeV in incoming photon energy. Due to the flat acceptance, covering the entire angular range,
and the well defined background subtraction, the full angular distributions and recoil polarization
distributions were accurately determined. A reliable normalization is obtained by comparing the
differential cross sections for the process
$\gamma p \rightarrow \eta p$ to the published data in this channel.
The statistical accuracy of our measurement is comparable to the two analyses (CLAS and SAPHIR) published
during the course of this work. The three analyses largely agree, both in shape and in absolute
magnitude. The systematic uncertainty of the measurements is 15 \%, mainly due to the uncertainty in the
normalization.

The present data are sensitive to inclusions of particular additional
resonances. The total cross section is well reproduced by both the Usov-Scholten and the Bonn-Gatchina models,
but inclusion of an additional resonance at about 1840 MeV is required by both models. In the Usov-Scholten model,
the new resonance has $P_{13}$ quantum numbers, in the Bonn-Gatchina model, a $P_{11}$ is preferred. Polarisation
experiments will be needed to come to final conclusions. Inclusion of the $S_{11}(1729)$ does not 
improve the agreement between the prediction of Usov and Scholten and measurement, neither for the differential cross 
sections, nor for the $\Sigma^+$ recoil polarization.

\begin{acknowledgement}
We thank the technical staff at ELSA and at all the participating
institutions for their important contributions to the success of the experiment. We gratefully acknowledge valuable discussions with Alexander Usov and Olaf Scholten. This work was performed as part of the research program of the Stichting voor Fundamenteel Onderzoek der Materie (FOM) with financial support from the Nederlandse Organisatie voor Wetenschappelijk Onderzoek (NWO).
We acknowledge financial support from the Deutsche Forschungsgemeinschaft
(DFG), the Schweizerischer Nationalfond and the Russian Foundation for Basic Research. U. Thoma thanks for an Emmy-Noether grant from
the DFG. A. V. Anisovich and A. V. Sarantsev acknowledge
support from the Alexander von Humboldt Foundation.

\end{acknowledgement}

\end{document}